\documentclass[10pt]{amsart}
\usepackage{amssymb}
\usepackage{enumerate}
\usepackage[dvips]{graphicx}
\DeclareGraphicsExtensions{.eps,.art,.ART,.ps}
\newcommand{\ba}{{\bf \hat{a}}}
\newcommand{\cM}{\mathcal{M}}
\newcommand{\cX}{\mathfrak{X}}
\newcommand{\soo}{\mathfrak{so}(3,1)}
\newcommand{\so}{\mathfrak{so}(3)}
\newcommand{\bsigma}{{\boldsymbol \sigma}}
\newcommand{\bxi}{{\boldsymbol \xi}}
\newcommand{\bbF}{\mathbb{F}}
\newcommand{\bbR}{\mathbb{R}}      

\newcommand{\tr}{\operatorname{tr}}

\newtheorem{Thm}{Theorem}[section]
\newtheorem{Prop}[Thm]{Proposition}

\newtheorem{Def}[Thm]{Definition}

\newtheorem{Remark}[Thm]{Remark}

\begin{document}
\begin{abstract}
We define the tangent Euler top in General Relativity through a constrained Lagrangian on the orthonormal frame bundle. The corresponding motions are studied to various degrees of approximation, the lowest of which is shown to yield the Mathisson-Papapetrou equations.
\end{abstract}
%
%
\title{Tangent Euler Top in General Relativity}
\author{Jos\'{e} Nat\'{a}rio}
\address{Centro de An\'alise Matem\'atica, Geometria e Sistemas Din\^amicos, Departamento de Matem\'atica, Instituto Superior T\'ecnico, 1049-001 Lisboa, Portugal}
\thanks{This work was partially supported by the Funda\c{c}\~ao para a Ci\^encia e a Tecnologia
through the Program POCI 2010/FEDER and by the grant POCI/MAT/58549/2004}
\maketitle
%
%
%
\section*{Introduction}
As is well known \cite{Born09, Herglotz10, Noether10, Rosen47, ST54, NJ59, PW63, Boyer65, WE66, WE67}, there are two separate sets of problems preventing the consideration of generic rigid bodies in General Relativity. The first is the lack of a global notion of simultaneity, which renders the notion of distance between two particles undefined. Associated to this is the inexistence of arbitrarily fast signals, which would be necessary for a rigid body to keep its exact shape. The second problem is that even if a family of spacelike slices is somehow singled out to define a notion of simultaneity, it is in general impossible to keep the mutual distances of more than four particles constant, due to the varying curvature of the slices.

Nevertheless, it is many times useful to consider approximate notions of rigidity in General Relativity \cite{D70, ER77, TG83, Voinov87, XTW04}. These are usually derived from multipole or post-Newtonian expansions, which can be quite formidable. A simpler method for obtaining the (approximate) motion of a small (approximately) rigid body from a finite-dimensional Lagrangian, mirroring what is done in Newtonian mechanics, would therefore be desirable.

A possibility (suggested by Turakulov \cite{Turakulov97} in the context of Newtonian mechanics on a curved space) is to approximate the motion of a rigid body by the motion of an orthonormal frame (determined by a naturally defined Lagrangian). This avoids the curvature problem. The simultaneity problem can be overcome by adding the natural constraint that the frame's timelike unit vector should be tangent to the motion of the base point (related ideas in flat spacetime can be found in \cite{HR74, Ellis82, Takabayasi82}).

The consequences of this model are explored in the present paper. Section 1 begins by briefly reviewing the Euler top, in order to set up the notation. The Newtonian tangent Euler top on a curved space, which is a toy model for its relativistic counterpart, is investigated in Section 2. This generalizes \cite{Turakulov97}, where the discussion was restricted to spherical tops. The relativistic tangent Euler top is defined in Section 3, as well as simpler Lagragians determining the spinning motion only. An approximation scheme for these is devised in Section 4, where the quadratic and cubic approximations are examined. The first (but not the second) is seen to lead to the usual Euler top motion with respect to a Fermi-Walker transported frame. Section 5 studies the exact spinning motion when the base point is moving along a geodesic. The corresponding Lagrangian is shown to yield a (new) completely integrable system on $SO(3)$. Finally, Section 6 analyzes the full Lagrangian in the quadratic approximation. It is found that the motion of the frame's base point is determined by the Mathisson-Papapetrou equation of motion subject to the Mathisson-Pirani spin supplementary condition \cite{Mathisson37,P51, Pirani56, MTW73, S99, Stephani04}.

The present paper thus succeeds in deriving the Mathisson-Papapetrou equation from a Lagrangian, defined on the (finite-dimensional) orthonormal frame bundle, which is the natural generalization of the Lagrangian for a free rigid body in Newtonian mechanics. This is quite different from other approaches inspired on variational principles \cite{Kunzle72, YB93, Porto06}, where the authors start with the Mathisson-Papapetrou equation and proceed to construct an action (generally with a much less clear physical meaning). Moreover, this Lagrangian formulation allows for the study of the rotational motion (which can be analyzed for arbitrary motions of the center of mass), instead of considering only the evolution of the angular momentum tensor. This leads to interesting new mechanical systems on $SO(3)$.

We use the conventions of \cite{MTW73} (including the Einstein summation convention), except for the meaning of the indices, which is the following: latin indices will always be associated to an orthonormal frame; they will be spacetime indices (ranging from $0$ to $3$) if they are from the beginning of the alphabet ($a,b,\ldots$), and space indices (ranging from $1$ to $3$) if the are from the middle of the alphabet ($i,j,\ldots$). Similarly, greek indices will always be associated to a coordinate system; they will be spacetime indices if the are from the beginning of the alphabet ($\alpha,\beta,\ldots$), and space indices if the are from the middle of the alphabet ($\mu,\nu,\ldots$)\footnote{This complication is unfortunate but inevitable, as we work in the orthonormal frame bundle and often distinguish between space and time.}. As is usual, we will not worry about the vertical position of space indices on orthonormal frames.
%
%
%
\section{Euler top}
We shall use this brief review of the Euler top to set up our notation. For simplicity we will consider only the case where the Euler top which spins about its center of mass.

\begin{Def}
The {\bf reference configuration} of an Euler top is a positive finite Borel measure $m$ on $\bbR^3$, not supported on any $1$-dimensional subspace, and such that
\[
\int_{\bbR^3} \bxi \, dm = {\bf 0}, \quad \quad \quad \int_{\bbR^3} \| \bxi \|^2 dm < + \infty.
\]
The {\bf mass} of the Euler top is $\cM = m(\bbR^3)$, and its {\bf Euler tensor} is the symmetric matrix
\[
I_{ij} = \int_{\bbR^3} \xi^i \xi^j dm.
\]
A {\bf motion} of the Euler top is an Euler-Lagrange curve for the Lagrangian $L:TSO(3)\to \bbR$ given by\footnote{Here we use the usual convention that $\dot{S}$ can either mean the time derivative of a curve $S:\bbR \to SO(3)$ or a tangent vector on the tangent bundle $TSO(3)$.}
\[
L(S,\dot{S}) = \frac12 \int_{\bbR^3} \langle \dot{S} \bxi, \dot{S} \bxi \rangle \, dm = \frac12 \dot{S}_{ij} I_{jk} \dot{S}_{ik} = \frac12 \tr(\dot{S} I \dot{S}^t).
\]
\end{Def}

As is well known \cite{Ar97, MR99}, the Euler top is completely integrable: in addition to the total energy $H=L$, one can obtain commuting first integrals from the conservation of angular momentum. In fact, the Euler-Lagrange equations themselves are equivalent to the conservation of angular momentum.

\begin{Thm}
If $S:\bbR \to SO(3)$ is a motion of the Euler top, then the {\bf angular momentum matrix}
\[
\sigma = S(IA+AI)S^t \in \so
\]
is constant, where $A: \bbR \to \so$ is such that $\dot{S}=SA$.
\end{Thm}

\begin{proof}
$SO(3)$ acts on itself by left multiplication, and $L=-\frac12 \tr(A I A)$ is clearly $SO(3)$-invariant. By Noether's theorem, the function $\bbF L(X^B)$ is conserved along the motion for all $B \in \so$, where $\bbF L$ is the fiber derivative of the Lagrangian and $X^B \in \cX(SO(3))$ is the infinitesimal action of $B \in \so$. Now
\[
(X^B)_S = \left.\frac{d}{dt}\right|_{t=0} \exp(tB) S = BS = SS^tBS,
\]
and hence
\[
\bbF L(X^B) = - \frac12 \tr(S^tBS I A) - \frac12 \tr(A I S^tBS) = - \frac12 \tr(S(IA+AI)S^tB).
\]
The formula $\langle\langle C,D \rangle\rangle = - \frac12 \tr(CD)$ defines an inner product on $\so$. Since $\langle\langle \sigma, B \rangle\rangle$ is conserved for all $B \in \so$, so is $\sigma$.
\end{proof}

\begin{Remark} \hspace{1cm}
\begin{enumerate}
\item
The matrix $\sigma$ is related to the {\bf angular momentum vector} $\bsigma$ through
\[
\sigma \bxi = \bsigma \times \bxi
\]
for all $\bxi \in \bbR^3$, i.e. $\sigma_{ik} = \varepsilon_{ijk}\sigma^j$.
\item
We have $\sigma = S \Sigma S^t$ with $\Sigma=IA+AI \in \so$. Conservation of $\sigma$ is equivalent to the {\bf Euler equation} \cite{Ar97, MR99, O02}
\[
\dot{\Sigma}=\Sigma A-A\Sigma.
\]
\end{enumerate}
\end{Remark}
%
%
%
\section{Tangent Euler top on a curved space}
Let $OM$ be the bundle of positive orthonormal frames on an oriented\footnote{The hypothesis of orientability is not crucial, as one can always pass to the orientable double cover.} $3$-dimensional Riemannian manifold $(M, \langle \cdot, \cdot \rangle)$. A curve $E:\bbR \to OM$ is given by
\[
E(t) = (c(t), E_1(t), E_2(t), E_3(t)),
\]
where $c:\bbR \to M$ is a curve on $M$ and $E_1,E_2,E_3:\bbR \to TM$ are vector fields along $c$ which form a positive orthonormal frame at $c(t)$ for each $t \in \bbR$.

\begin{Def} \label{DefmotionEulertop}
 A {\bf motion} of a tangent Euler top on $(M, \langle \cdot, \cdot \rangle)$ is an Euler-Lagrange curve for the Lagrangian $L:TOM \to \bbR$ given by
\[
L(E,\dot{E}) = \frac12 \int_{\bbR^3} \left\langle \dot{c} + \xi^i \nabla_{\dot{c}} E_i, \dot{c} + \xi^j \nabla_{\dot{c}} E_j \right\rangle \, dm,
\]
where $\nabla$ is the Levi-Civita connection and $m$ is the reference configuration of an Euler top.
\end{Def}

\pagebreak

\begin{Remark} \hspace{1cm}
\begin{enumerate}
\item
Notice that if $(M, \langle \cdot, \cdot \rangle)$ is $\bbR^3$ with the Euclidean metric then this Lagrangian is just the Lagrangian for a free rigid body. 
\item
Physically, one can expect a tangent Euler top to represent the motion of an approximately rigid extended body in the limit in which its size is much smaller than the local radius of curvature. 
\end{enumerate}
\end{Remark}

\begin{Thm} \label{ThmmotionEulertop}
The motions of a tangent Euler top on $(M, \langle \cdot, \cdot \rangle)$ are as follows: the frame $\{E_1,E_2,E_3\}$ rotates exactly as an Euler top with respect to any frame $\{\hat{E}_1, \hat{E}_2, \hat{E}_3 \}$ which is parallel-transported along the motion of the base point; the base point moves according to the equation\footnote{As is usual, $\frac{D}{dt}=\nabla_{\dot{c}}$ is the covariant derivative along the curve $c:\bbR \to M$.}
\begin{equation} \label{NewtonMathisson}
\cM \, \frac{D\dot{c}}{dt} + \frac12 \left( \iota(\dot{c})\hat{\Omega}_{ij} \right)^\sharp \sigma_{ij} = 0,
\end{equation}
where $\hat{\Omega}_{ij}$ are the curvature forms for the frame $\{\hat{E}_1, \hat{E}_2, \hat{E}_3 \}$, $^\sharp:T^*M \to TM$ is the isomorphism induced by the metric and $\sigma \in \so$ is the angular momentum matrix.
\end{Thm}

\begin{proof}
We define a local trivialization $OM|_U \cong U \times SO(3)$ by choosing a local orthonormal frame $\{\hat{E}_1, \hat{E}_2, \hat{E}_3 \} \subset \cX(U)$ on a sufficiently small open set $U \subset M$. For this trivialization we have
\[
E_i(t) = S_{ji}(t) (\hat{E}_j)_{c(t)} \Rightarrow \nabla_{\dot{c}} E_i = \dot{S}_{ji} \hat{E}_j + S_{ji} \hat{\omega}_{\,\,\, j}^k(\dot{c}) \hat{E}_k,
\]
where $\hat{\omega}_{\,\,\, i}^j$ are the connection forms associated to our local frame. The Lagrangian can therefore be written as
\[
L = T + K + C + F
\]
where
\begin{align*}
& T = \frac12 \int_{\bbR^3} \langle \dot{c}, \dot{c} \rangle \, dm = \frac12 \cM \, \langle \dot{c}, \dot{c} \rangle; \\
& K = \frac12 \int_{\bbR^3} \xi^i \dot{S}_{ki} \xi^j \dot{S}_{lj} \langle \hat{E}_k, \hat{E}_l \rangle \, dm = \frac12 \tr(\dot{S}I\dot{S}^t); \\
& C = \int_{\bbR^3} \xi^i \dot{S}_{ki} \xi^j S_{lj} \hat{\omega}_{\,\,\,\, l}^m(\dot{c}) \langle \hat{E}_k, \hat{E}_m \rangle \, dm = \tr(\dot{S}IS^t\hat{\omega}^t(\dot{c})); \\
& F = \frac12 \int_{\bbR^3} \xi^i S_{ki} \hat{\omega}_{\,\,\, k}^l(\dot{c})\xi^j S_{mj} \hat{\omega}_{\,\,\, m}^n(\dot{c}) \langle \hat{E}_l, \hat{E}_n \rangle \, dm = \frac12 \tr(\hat{\omega}(\dot{c})SIS^t\hat{\omega}^t(\dot{c})).
\end{align*}
Our task is now to write the Euler-Lagrange equations. For the $SO(3)$ part, we notice that if the frame $\{\hat{E}_1, \hat{E}_2, \hat{E}_3 \}$ happens to be parallel transported along $c$ then $\hat{\omega}(\dot{c})=0$, and hence only terms coming from the rotational kinetic energy $K$ survive in the Euler-Lagrange equations. Therefore $\{E_1, E_2, E_3\}$ rotates with respect to $\{\hat{E}_1, \hat{E}_2, \hat{E}_3 \}$ exactly as an Euler top with reference configuration $m$. For the motion of the base point, there will be terms coming from both the translational kinetic energy $T$ and the Coriolis term $C$ (but not from the frame term $F$, which is quadratic in $\hat{\omega}(\dot{c})$). As is well known,
\[
\frac{d}{dt}\left( \frac{\partial T}{\partial \dot{x}^\mu} \right) - \frac{\partial T}{\partial x^\mu} = \cM \, g_{\mu \nu} \frac{D \dot{x}^\nu}{dt},
\]
where $g_{\mu \nu}$ is the matrix of the metric. Since the trace of the product of a symmetric and an anti-symmetric matrices is zero, we can write
\[
C = \tr(SAIS^t\hat{\omega}^t(\dot{c})) = \frac12 \tr(S(AI-IA^t)S^t\hat{\omega}^t(\dot{c})) = \frac12 \tr(\sigma\hat{\omega}^t(\dot{c})) = \frac12 \sigma_{ij} \hat{\omega}_{\,\,\, j\mu}^i \dot{x}^\mu.
\]
Therefore
\[
\frac{d}{dt}\left( \frac{\partial C}{\partial \dot{x}^\mu} \right) - \frac{\partial C}{\partial x^\mu} = \frac12 \sigma_{ij} (\partial_\nu\hat{\omega}_{\,\,\, j\mu}^i - \partial_\mu\hat{\omega}_{\,\,\, j\nu}^i) \dot{x}^\nu = \frac12 \sigma_{ij} \hat{\Omega}_{\,\,\, j\nu\mu}^i\dot{x}^\nu,
\]
where $\hat{\Omega}_{\,\,\, j}^i = d\hat{\omega}_{\,\,\, j}^i + \hat{\omega}_{\,\,\, k}^i \wedge \hat{\omega}_{\,\,\, j}^k$ are the curvature forms associated to $\{\hat{E}_1, \hat{E}_2, \hat{E}_3 \}$. Consequently, the equations of motion for the base point are
\[
\cM \, \frac{D \dot{x}^\mu}{dt} + \frac12 \sigma_{ij}\hat{\Omega}_{ij\nu}^{\,\,\,\,\,\,\,\,\,\mu}\dot{x}^\nu = 0.
\]
\end{proof}

\begin{Remark} \hspace{1cm}
\begin{enumerate}
\item
If $(M, \langle \cdot, \cdot \rangle)$ is $\bbR^3$ with the Euclidean metric then the curvature forms vanish and we obtain the familiar result that the center of mass of a free rigid body moves with constant velocity.
\item
We can write Equation~(\ref{NewtonMathisson}) as
\[
\cM \frac{D \dot{x}^\mu}{dt} = \frac12 R^\mu_{\,\,\, \nu\kappa\lambda} \dot{x}^\nu s^{\kappa \lambda},
\]
where
\[
s = \sigma_{ij} \hat{E}_i \otimes \hat{E}_j
\]
is the {\bf angular momentum tensor}. As noticed by Turakulov \cite{Turakulov97}, this equation is similar to the Mathisson-Papapetrou equation for a spinning particle in General Relativity \cite{Mathisson37, P51}. The fact that the angular momentum matrix is constant on the parallel transported frame $\{\hat{E}_1, \hat{E}_2, \hat{E}_3 \}$ is translated into the fact that the angular momentum tensor satisfies the parallel transport equation
\[
\frac{Ds^{\mu\nu}}{dt}=0.
\]
\end{enumerate}
\end{Remark}
%
%
%
\section{Tangent Euler top in General Relativity}
Let $OM$ be the bundle of positive future-pointing orthonormal frames on an oriented, time-oriented\footnote{Again this is not a crucial hypothesis, as one can always pass to the time-orientable double cover.} $4$-dimensional Lorentzian manifold $(M, \langle \cdot, \cdot \rangle)$. An {\bf admissible curve} $E:\bbR \to OM$ is given by
\[
E(t) = (c(t), E_0(t), E_1(t), E_2(t), E_3(t)),
\]
where $c:\bbR \to M$ is a future-directed timelike curve on $M$ parameterized by its proper time and $E_0,E_1,E_2,E_3:\bbR \to TM$ are vector fields along $c$ which form a positive orthonormal frame at $c(t)$ satisfying $E_0(t)=\dot{c}(t)$ for each $t \in \bbR$.

\begin{Def} \label{DefmotionEulertopGR}
 A {\bf motion} of a tangent Euler top on $(M, \langle \cdot, \cdot \rangle)$ is a (constrained) Euler-Lagrange curve for the Lagrangian $L:TOM \to \bbR$ given by
\[
L(E,\dot{E}) = \int_{\bbR^3} \left| \left\langle \dot{c} + \xi^i \nabla_{\dot{c}} E_i, \dot{c} + \xi^j \nabla_{\dot{c}} E_j \right\rangle\right|^\frac12 \, dm,
\]
where $\nabla$ is the Levi-Civita connection and $m$ is the reference configuration of an Euler top, chosen among all admissible curves, with unspecified final time.
\end{Def}

\begin{Remark} \hspace{1cm}
\begin{enumerate}
\item
This is the natural generalization of the Lagrangian for a (Newtonian) tangent Euler top on a curved space: one simply replaces the squared velocity by its relativistic length.
\item
Restricting the action to admissible curves means considering bodies which are rigid as seen by an observer comoving with the base point.
\item
We must leave the final time unspecified because we insisted that admissible curves should be parameterized by the proper time of the base point, which needs not be the same for all admissible curves with given endpoints.
\item
A tangent Euler top in General Relativity can therefore be physically interpreted as the motion of an extended body which is approximately rigid as seen by an observer placed at its center of mass, in the limit in which the body's size is much smaller than the local radius of curvature. 
\end{enumerate}
\end{Remark}

A simpler problem is obtained by focussing on the spinning motion. To do this, we regard the motion $c:\bbR \to M$ of the base point as given (which may or may not be the one arising from Definition~\ref{DefmotionEulertopGR}) and look for the positive orthonormal frame $\{ \dot{c}(t), E_1(t), E_2(t), E_3(t) \}$ along $c$ which maximizes the action. Choosing a positive orthonormal Fermi-Walker transported frame $\{ \dot{c}(t), \hat{E}_1(t), \hat{E}_2(t), \hat{E}_3(t) \}$ along $c$, so that
\[
\nabla_{\dot{c}} \dot{c}(t) = \hat{a}_i(t) \hat{E}_i(t), \quad \quad \quad \nabla_{\dot{c}} \hat{E}_i(t) = \hat{a}_i(t) \dot{c}(t),
\]
we have
\[
E_i(t) = S_{ji}(t) (\hat{E}_j)_{c(t)} \Rightarrow \nabla_{\dot{c}} E_i = \dot{S}_{ji} \hat{E}_j + S_{ji} a_j \dot{c}
\]
for some curve $S:\bbR \to SO(3)$. Therefore
\[
\left\langle \dot{c} + \xi^i \nabla_{\dot{c}} E_i, \dot{c} + \xi^j \nabla_{\dot{c}} E_j \right\rangle = -1 + \dot{S}_{ki} \xi^i \dot{S}_{kj} \xi^j - \hat{a}_k S_{ki} \xi^i \hat{a}_l S_{lj} \xi^j - 2 \hat{a}_j S_{ji} \xi^i.
\]

\begin{Def} \label{DefspinningmotionEulertopGR}
 A {\bf spinning motion} of a tangent Euler top on $(M, \langle \cdot, \cdot \rangle)$ is an (unconstrained) Euler-Lagrange curve for the Lagrangian $L:TSO(3) \times \bbR \to \bbR$ given by
\[
L(S,\dot{S},t) = \int_{\bbR^3} \left| 1 - \langle \dot{S} \bxi, \dot{S} \bxi \rangle + \langle S \bxi, \ba(t) \rangle^2 + 2 \langle S \bxi, \ba(t) \rangle \right|^\frac12 \, dm,
\]
where $m$ is the reference configuration of an Euler top and $\ba:\bbR \to \bbR^3$ represents the proper acceleration on a Fermi-Walker transported frame.
\end{Def}
%
%
%
\section{Approximate spinning motions} \label{section4}
Expanding the integrand in the Lagrangian for a spinning motion in power series of $\bxi$ and integrating we can write
\[
L = \cM + L^{(2)} + L^{(3)} + \ldots,
\]
where $L^{(n)}$ is the integral of the term of order $n$ in $\bxi$.

\begin{Def} \label{DefApprox}
An {\bf approximate spinning motion of order $n$} of a tangent Euler top is an Euler-Lagrange curve of the truncated Lagrangian
\[
L = \cM + L^{(2)} + \ldots + L^{(n)}.
\]
\end{Def}

\begin{Thm}
The approximate spinning motions of order $2$ of a tangent Euler top are exactly the motions of an Euler top with the same reference configuration.
\end{Thm}

\begin{proof}
One just needs to check that
\[
L^{(2)} = - \frac12 \int_{\bbR^3} \langle \dot{S} \bxi, \dot{S} \bxi \rangle \, dm.
\]
\end{proof}

\begin{Remark}
Notice that the proper acceleration $\ba$ does not appear in the expression of $L^{(2)}$: in this approximation, the spinning motion completely decouples from the translational motion. This is similar to what happens in the Newtonian case (and is, of course, the principle underlying the operation of a gyroscope).
\end{Remark}

\begin{Prop}
Generically, the approximate spinning motions of order $3$ of a tangent Euler top are {\bf not} the motions of an Euler top with the same reference configuration.
\end{Prop}

\begin{proof}
One just needs to check that
\[
L^{(3)} = \frac12 \int_{\bbR^3} \langle \dot{S} \bxi, \dot{S} \bxi \rangle \langle S \bxi, \ba \rangle \, dm,
\]
which generically is not a total time derivative.
\end{proof}

\begin{Remark}
$L^{(3)}$ can be thought of as (minus) the potential energy corresponding to the weight of the rotational kinetic energy in the constant ``gravitational field'' $-\ba$ induced by the acceleration of the center of mass. 
\end{Remark}
%
%
%
\section{Spinning motions along geodesics}
If the spinning motion is along a timelike geodesic, we have $\ba = {\bf 0}$, and the Lagrangian becomes $SO(3)$-invariant. In this case, we can apply Noether's Theorem.

\begin{Thm}
If $S:\bbR \to SO(3)$ is a spinning motion of a tangent Euler top along a timelike geodesic of $(M, \langle \cdot, \cdot \rangle)$, then $\sigma = S \Sigma S^t \in \so$ is constant, where
\[
\Sigma = \int_{\bbR^3} \frac{(A \bxi) \bxi^t - \bxi (A \bxi)^t}{| 1 - \langle A \bxi, A \bxi \rangle |^\frac12} \, dm \in \so
\]
and $A: \bbR \to \so$ is such that $\dot{S}=SA$.
\end{Thm}

\begin{proof}
We can write the Lagrangian as
\[
L(S, \dot{S}) = \int_{\bbR^3} \left| 1 - \tr \left((\dot{S}\bxi) (\dot{S} \bxi)^t\right) \right|^\frac12 \, dm =
\int_{\bbR^3} \left| 1 - \tr \left( A \bxi \bxi^t A^t\right) \right|^\frac12 \, dm.
\]
By Noether's theorem, $\bbF L(X^B)$ is conserved along the motion for all $B \in \so$, where $X^B \in \cX(SO(3))$ is the infinitesimal action of $B$. Since
\begin{align*}
\bbF L(X^B) & = - \frac12 \int_{\bbR^3} \frac{\tr \left( S^t B S \bxi \bxi^t A^t + A \bxi \bxi^t S^t B^t S \right)}{\left| 1 - \tr \left( A \bxi \bxi^t A^t\right) \right|^\frac12} \, dm \\
& = \frac12 \tr \left( B S \left(\int_{\bbR^3} \frac{A \bxi \bxi^t - \bxi \bxi^t A^t}{\left| 1 - \langle A \bxi, A \bxi \rangle \right|^\frac12} \, dm \right) S^t \right),
\end{align*}
we see that $-\frac12 \tr(B\sigma) = \langle\langle B, \sigma \rangle\rangle$ is conserved for all $B \in \so$, and so is $\sigma$.
\end{proof}

\begin{Remark} \hspace{1cm}
\begin{enumerate}
\item
The angular momentum vector $\bsigma$, related to the matrix $\sigma$ through
\[
\sigma \bxi = \bsigma \times \bxi
\]
for all $\bxi \in \bbR^3$, can be seen to have the familiar expression
\[
\bsigma = \int_{\bbR^3} \frac{(S \bxi) \times (\dot{S} \bxi)}{| 1 - \langle \dot{S} \bxi, \dot{S} \bxi \rangle |^\frac12} \, dm.
\]
\item
This system is completely integrable: in addition to the total energy 
\[
K = -H = - \bbF L(A) + L = \int_{\bbR^3} \frac{1}{\left| 1 - \langle A \bxi, A \bxi \rangle \right|^\frac12} \, dm,
\]
one can obtain commuting first integrals from the conservation of angular momentum. In fact, the Euler-Lagrange equations themselves are equivalent to the conservation of angular momentum. The images of the motions $A:\bbR \to \so$ can be obtained from the intersections of the level surfaces of the functions $K(A)$ and $\langle\langle \Sigma(A), \Sigma(A) \rangle\rangle = -\frac12\tr(\Sigma(A)^2)$.
\item
We have
\[
L(A) = \cM - \frac12 I_{ij} A_{ki} A_{kj} - \frac18 I_{ijkl} A_{mi} A_{mj} A_{nk} A_{nl} -  \ldots
\]
where
\[
I_{ijkl} = \int_{\bbR^3} \xi^i \xi^j \xi^k \xi^l dm, \quad \ldots
\]
Therefore the Euler tensor is not enough to characterize a relativistic tangent Euler top: the higher multipoles are also required (and become more important as the spinning motion becomes more relativistic).
\end{enumerate}
\end{Remark}
%
%
%
\section{Quadratically approximate motions}
We can extend Definition~\ref{DefApprox} to the general motion of a tangent Euler top.

\begin{Def}
 A {\bf quadratically approximated motion} of a tangent Euler top on $(M, \langle \cdot, \cdot \rangle)$ is an Euler-Lagrange curve for the Lagrangian $L:TOM \to \bbR$ given by
\[
L(E,\dot{E}) = \cM - \frac12 I_{ij} \langle \nabla_{\dot{c}}E_i, E_k \rangle \langle \nabla_{\dot{c}}E_j, E_k \rangle,
\]
chosen among all admissible curves, with unspecified final time.
\end{Def}

\begin{Thm} 
Quadratically approximated motions of a tangent Euler top on $(M, \langle \cdot, \cdot \rangle)$ are as follows: $\{E_1,E_2,E_3\}$ rotates exactly as an Euler top with respect to $\{\hat{E}_1, \hat{E}_2, \hat{E}_3 \}$, where $\{\hat{E}_0, \hat{E}_1, \hat{E}_2, \hat{E}_3 \}: \bbR \to OM$ is Fermi-Walker transported along the motion of the base point and satisfies $\hat{E}_0=\dot{c}$; the base point moves according to the equation
\begin{equation} \label{Mathisson}
(\cM + K) \, \frac{D\dot{c}}{dt} + \frac12 \left( \iota(\dot{c})\hat{\Omega}_{ij} \right)^\sharp \sigma_{ij} - \hat{E}_i \sigma_{ij} \frac{d}{dt} \left\langle \frac{D\dot{c}}{dt}, \hat{E}_j \right\rangle = 0,
\end{equation}
where $K$ is the rotational kinetic energy, $\hat{\Omega}_{ab}$ are the curvature forms for the frame $\{\hat{E}_0, \hat{E}_1, \hat{E}_2, \hat{E}_3 \}$, $^\sharp:T^*M \to TM$ is the isomorphism induced by the metric and $\sigma \in \so$ is the angular momentum matrix.
\end{Thm}

\begin{proof}
The spinning motion with respect to a Fermi-Walker transported frame was already shown to be that of an Euler top in Section~\ref{section4}.

To compute the motion of the base point we define a local trivialization $OM|_U \cong U \times SO^\uparrow(3,1)$ by choosing a local orthonormal frame $\{\hat{E}_0, \hat{E}_1, \hat{E}_2, \hat{E}_3 \} \subset \cX(U)$ on a sufficiently small open set $U \subset M$. For this trivialization we have
\[
E_i(t) = S^a_{\,\,\,i}(t) (\hat{E}_a)_{c(t)} \Rightarrow \nabla_{\dot{c}} E_i = \dot{S}^{a}_{\,\,\, i} \hat{E}_a +  S^a_{\,\,\,i} \hat{\omega}_{\,\,\, a}^b(\dot{c}) \hat{E}_b,
\]
where $\hat{\omega}_{\,\,\, a}^b$ are the connection forms associated to our local frame, and hence the Lagrangian can be written as
\[
L = \cM - \frac12 I_{ij} \left(\hat{\omega}_{ab}(\dot{c})S^a_{\,\,\,k}S^b_{\,\,\,i}+A_{ki}\right)\left(\hat{\omega}_{cd}(\dot{c})S^c_{\,\,\,k}S^d_{\,\,\,j}+A_{kj}\right),
\]
where $A=(A^a_{\,\,\,b}) \in \soo$ is defined by $\dot{S}=SA \Leftrightarrow \dot{S}^a_{\,\,\,b} = S^a_{\,\,\,c} A^c_{\,\,\,b}$. Notice that the conditions $S \in O(3,1)$ and $A \in \soo$ are conveniently written by raising and lowering indices as $(S^{-1})^a_{\,\,\,b}=S_b^{\,\,\,a}$ and $A_{ab}=-A_{ba}$.

The best way to deal with the constraint while avoiding introducing coordinates on $SO^\uparrow(3,1)$ is to to set up our problem as an optimal control problem (see for instance \cite{Lawden63}, \cite{Pinch93}). As state variables we choose the local coordinates $x^\alpha$ of the base point, the components $S_a^{\,\,\,b}$ of the frame vectors and the action $\theta$. As control variables we simply chose $6$ independent components of the anti-symmetric matrix $(A_{ab})$. The state variables depend on the control variables through the solution of the Cauchy problem
\[
\begin{cases}
\dot{x}^\alpha = \hat{E}^\alpha_a S^a_{\,\,\, 0} \\
\dot{S}^a_{\,\,\,b} = S^a_{\,\,\,c} A^c_{\,\,\,b} \\
\dot{\theta} = \cM - \frac12 I_{ij} \left(\hat{\omega}_{ab\alpha}\hat{E}^\alpha_c S^c_{\,\,\,0} S^a_{\,\,\,i}S^b_{\,\,\,k}+A_{ik}\right)\left(\hat{\omega}_{de\alpha}\hat{E}^\alpha_f S^f_{\,\,\,0} S^d_{\,\,\,j}S^e_{\,\,\,k}+A_{jk}\right)
\end{cases}
\]
Our objective is to maximize $\theta_1 = \theta(t_1)$, i.e.~minimize $\Phi=-\theta_1$, where the final time $t_1$ is not fixed. To do so, we introduce the multipliers $\lambda_\alpha, \Lambda_a^{\,\,\, b}, \lambda$ (one for each differential equation) and set up the Hamiltonian function
\begin{align*}
H & = \lambda_\alpha \hat{E}^\alpha_a S^a_{\,\,\, 0} + \Lambda_a^{\,\,\, b} S^a_{\,\,\,c} A^c_{\,\,\,b} \\
& + \lambda \left[ \cM - \frac12 I_{ij} \left(\hat{\omega}_{ab\alpha}\hat{E}^\alpha_c S^c_{\,\,\,0} S^a_{\,\,\,i}S^b_{\,\,\,k}+A_{ik}\right)\left(\hat{\omega}_{de\beta}\hat{E}^\beta_f S^f_{\,\,\,0} S^d_{\,\,\,j}S^e_{\,\,\,k}+A_{jk}\right)\right].
\end{align*}
According to the Pontryagin Maximum Principle, we must select the values of the control variables which maximize $H$. The minimizing curve will then be obtained by solving the Hamilton equations
\[
\begin{cases}
\dot{\lambda}_\alpha = - \frac{\partial H}{\partial x^\alpha} \\
\dot{\Lambda}_a^{\,\,\, b} = - \frac{\partial H}{\partial S^a_{\,\,\,b}}\\
\dot{\lambda} = - \frac{\partial H}{\partial \theta}
\end{cases}
\]
together with the Cauchy problem. The multiplier $\lambda$ is clearly constant; since $\theta_1$ is not specified, we must select
\[
\lambda(t_1) = - \frac{\partial \Phi}{\partial \theta_1} = 1
\]
(hence $\lambda \equiv 1$). Moreover, $H$ is constant along the solution. Since $t_1$ is not fixed, we must have $H \equiv 0$.

Because we already know what the spinning motion will be, we do not have to write out all the minimum equations. Also, we shall assume that the trivializing frame $\{\hat{E}_0, \hat{E}_1, \hat{E}_2, \hat{E}_3 \}$ happens to be Fermi-Walker transported along the minimizing curve, with $\hat{E}_0=\dot{c}$. As a consequence, we will have
\[
\hat{\omega}_{ij\alpha}\hat{E}_0^{\alpha} = 0, \quad \quad \hat{\omega}_{i0\alpha}\hat{E}_0^{\alpha} = \hat{a}_i, \quad \quad S^a_{\,\,\, 0} = \delta^a_{\,\,\, 0}
\]
along this curve. In particular,
\[
\dot{S}^a_{\,\,\, 0} = 0 \Leftrightarrow S^a_{\,\,\, c} A^c_{\,\,\, 0} = 0 \Leftrightarrow A^a_{\,\,\, 0} = 0 \Leftrightarrow A_{a0} = - A_{0a} = 0
\]
along the minimizing curve.

We now write the relevant minimum conditions. Maximizing $H$ with respect to the control variables $A_{0i}$ yields
\begin{equation} \label{Pontryagin1}
\frac{\partial H}{\partial A_{0i}} = 0 \Leftrightarrow \Lambda_a^{\,\,\, i} S^{a0} - \Lambda_a^{\,\,\, 0} S^{ai} = 0 \Leftrightarrow \Lambda^{0i} - \Lambda_a^{\,\,\, 0} S^{ai} = 0.
\end{equation}
We have the Hamilton equations
\begin{align} \label{Hamilton1}
\dot{\Lambda}_l^{\,\,\, 0} = - \frac{\partial H}{\partial S^l_{\,\,\,0}} & = - \lambda_\alpha \hat{E}^\alpha_l + \Lambda_l^{\,\,\, b} A^0_{\,\,\,b} \\ \nonumber
& \quad + I_{ij} \left(\hat{\omega}_{ab\alpha}\hat{E}^\alpha_l S^a_{\,\,\,i}S^b_{\,\,\,k}\right)\left(\hat{\omega}_{de\beta}\hat{E}^\beta_f S^f_{\,\,\,0} S^d_{\,\,\,j}S^e_{\,\,\,k}+A_{jk}\right) \\ \nonumber
& = - \lambda_\alpha \hat{E}^\alpha_l
\end{align}
and
\begin{align} \label{Hamilton2}
\dot{\Lambda}_0^{\,\,\, l} = - \frac{\partial H}{\partial S^0_{\,\,\,l}} & = - \Lambda_0^{\,\,\, b} A^l_{\,\,\,b} + I_{lj} \left(\hat{\omega}_{0b\alpha}\hat{E}^\alpha_c S^c_{\,\,\,0} S^b_{\,\,\,k}\right)\left(\hat{\omega}_{de\beta}\hat{E}^\beta_f S^f_{\,\,\,0} S^d_{\,\,\,j}S^e_{\,\,\,k}+A_{jk}\right) \\ \nonumber
& \quad + I_{ij} \left(\hat{\omega}_{a0\alpha}\hat{E}^\alpha_c S^c_{\,\,\,0} S^a_{\,\,\,i}\right)\left(\hat{\omega}_{de\beta}\hat{E}^\beta_f S^f_{\,\,\,0} S^d_{\,\,\,j}S^e_{\,\,\,l}+A_{jl}\right) \\ \nonumber
& = - \Lambda_0^{\,\,\, i} A_{li} - I_{lj} \hat{a}_m S^m_{\,\,\, k} A_{jk} + I_{ij} \hat{a}_m S^m_{\,\,\, i} A_{jl}.
\end{align}
Differentiating (\ref{Pontryagin1}) with respect to $t$ and using (\ref{Hamilton1}), (\ref{Hamilton2}), and again (\ref{Pontryagin1}) yields
\[
\lambda_\alpha \hat{E}^\alpha_j S_{ji} = - (I_{ij}A_{jk} + A_{ij}I_{jk}) S_{mk} \hat{a}_m,
\]
from which one readily obtains
\begin{equation} \label{Final1}
\lambda_\alpha \hat{E}^\alpha_l = - \sigma_{lm} \hat{a}_m.
\end{equation}
We must also consider the Hamilton equation
\begin{align} \label{Hamilton3}
\dot{\lambda}_\mu = - \frac{\partial H}{\partial x^\mu} & = - \lambda_\alpha \partial_\mu\hat{E}^\alpha_0 + I_{ij} \left(\partial_\mu\hat{\omega}_{ab\alpha}\hat{E}^\alpha_0 S^a_{\,\,\,i}S^b_{\,\,\,k}\right) A_{jk} \\
\nonumber & \quad + I_{ij} \left(\hat{\omega}_{ab\alpha}\partial_\mu\hat{E}^\alpha_0 S^a_{\,\,\,i}S^b_{\,\,\,k}\right) A_{jk} 
\end{align}
We now use our freedom in the choice of local coordinates to select Fermi normal coordinates $(t,x^1,x^2,x^3)$, defined by the parameterization
\[
\varphi(t,x^1,x^2,x^3) = \exp_{c(t)} (x^1 \hat{E}_1 + x^2 \hat{E}_2 + x^3 \hat{E}_3),
\]
where $\exp_p:T_pM \to M$ is the geodesic exponential map. To first order in $x$ we can then choose\footnote{Here we use latin indices $i,j,\ldots$ for the coordinates, as they are associated to the Fermi-Walker transported frame. By the same token, the coordinate $t$ coincides with the proper time $t$ along the minimizing curve.}
\[
\hat{E}_0 = (1-\hat{a}_i x^i)  \frac{\partial}{\partial t}, \quad \quad \quad \hat{E}_i = \frac{\partial}{\partial x^i}
\]
(see \cite{Poisson04}). In particular,  $\nabla_{\hat{E}_i}\hat{E}_j=0$ for $x=0$, from where we can deduce that $\hat{\omega}_{ij}=0$ and $\hat{\omega}_{0i}(\hat{E}_j)=0$ along the minimizing curve. Equation (\ref{Hamilton3}) then becomes
\begin{equation} \label{Final2}
\dot{\lambda}_n = \lambda_t \hat{a}_n + I_{ij} \left(\partial_n\hat{\omega}_{lm\alpha}\hat{E}^\alpha_0 S_{li}S_{mk}\right) A_{jk}.
\end{equation}
The quantity $\lambda_t$ in this equation can be obtained from
\begin{equation} \label{Hamilton4}
H = 0 \Leftrightarrow \lambda_t + \Lambda_{kj} S_{ki} A_{ij} + \cM - \frac12 I_{ij} A_{ik} A_{jk} = 0.
\end{equation}
Maximizing $H$ with respect to $A_{ij}$ (with $i<j$, say) yields further relations:
\begin{align*}
\frac{\partial H}{\partial A_{ij}} = 0 & \Leftrightarrow \Lambda_a^{\,\,\, j} S^{ai} - \Lambda_a^{\,\,\, i} S^{aj} - I_{ik}A_{kj} + I_{jk} A_{ki} = 0 \\
& \Leftrightarrow \Lambda_{kj} S_{ki} - \Lambda_{ki} S_{kj} = I_{ik}A_{kj} - I_{jk} A_{ki}. 
\end{align*}
These can be used to rewrite the second term in (\ref{Hamilton4}), which is
\[
\tr(SA\Lambda^t) = \frac12\tr(SA\Lambda^t+\Lambda A^tS^t) = \frac12\tr(\Lambda^tSA-S^t\Lambda A) = \frac12\tr((\Lambda^tS-S^t\Lambda) A).
\]
After the substitution we obtain
\[
\lambda_t = - \cM - \frac12 I_{ij} A_{ik} A_{jk} = - (\cM + K),
\]
where $K$ is the rotational kinetic energy of the Euler top.

Because $\partial_n\hat{\omega}_{lm\alpha}\hat{E}^\alpha_0$ is anti-symmetric in $l,m$, the quantity multiplying it in (\ref{Final2}) can be replaced by
\[
\frac12 ( I_{ij} S_{li}S_{mk} A_{jk} - I_{ij} S_{mi}S_{lk} A_{jk} ) = \frac12 S_{li} (I_{ij} A_{jk} - I_{kj} A_{ji}) S_{mk} = \frac12 \sigma_{lm}.
\]
Consequently, Equation~(\ref{Final2}) can be written as
\begin{equation} \label{Final3}
\dot{\lambda}_n = - (\cM + K) \hat{a}_n + \frac12 \partial_n\hat{\omega}_{lm\alpha}\hat{E}^\alpha_0 \sigma_{lm}.
\end{equation}

The curvature forms are given by
\[
\hat{\Omega}_{\,\,\, a}^b=d\hat{\omega}_{\,\,\, a}^b + \hat{\omega}_{\,\,\, c}^b \wedge \hat{\omega}_{\,\,\, a}^c \Rightarrow \hat{\Omega}_{ij} = d\hat{\omega}_{ij} - \hat{\omega}_{i0} \wedge \hat{\omega}_{0j} + \hat{\omega}_{ik} \wedge \hat{\omega}_{kj}.
\]
In components,
\[
\hat{\Omega}_{ijnt} = \partial_n\hat{\omega}_{ijt} - \partial_t\hat{\omega}_{ijn} - \hat{\omega}_{i0n} \hat{\omega}_{0jt} + \hat{\omega}_{i0t} \hat{\omega}_{0jn} + \hat{\omega}_{ikn} \hat{\omega}_{kjt} - \hat{\omega}_{ikt} \hat{\omega}_{kjn}.
\]
Recalling that with our choices $\hat{\omega}_{ij}=0$ and $\hat{\omega}_{0i}(\hat{E}_j)=0$ along the minimizing curve, we have
\[
\hat{\Omega}_{ijnt} = \partial_n\hat{\omega}_{ijt}
\]
on this curve. Therefore (\ref{Final3}) is equivalent to
\[
\dot{\lambda}_n = - (\cM + K) \hat{a}_n + \frac12 \hat{\Omega}_{lmnt} \sigma_{lm}.
\]
On the other hand, (\ref{Final1}) can be written as
\[
\lambda_l =  -\sigma_{lm} \hat{a}_m.
\]
From these two equations we finally deduce
\[
(\cM + K) \hat{a}_i - \frac12 \hat{\Omega}_{jkit} \sigma_{jk} - \sigma_{ij} \frac{d\hat{a}_j}{dt} = 0,
\]
which are the nontrivial components of Equation~(\ref{Mathisson}).
\end{proof}

\begin{Remark} \hspace{1cm}
\begin{enumerate}
\item
As one would expect, the remaining minimum conditions lead to the Euler equation.
\item
We can write Equation~(\ref{Mathisson}) as
\[
(\cM + K) \frac{D \dot{x}^\alpha}{dt} = \frac12 R^\alpha_{\,\,\, \beta\gamma\delta} \dot{x}^\beta s^{\gamma \delta} + s^\alpha_{\,\,\, \beta} \frac{D^2 \dot{x}^\beta}{dt^2},
\]
where
\[
s = \sigma_{ij} \hat{E}_i \otimes \hat{E}_j
\]
is the {\bf angular momentum tensor}\footnote{Notice however that many authors define the angular momentum tensor to be $\widetilde{s}=-s$, corresponding to the alternative relation $\widetilde{\sigma}_{ij} = \varepsilon_{ijk}\sigma^k$ between the angular momentum matrix and the angular momentum vector.}. This equation is exactly the Mathisson-Papapetrou equation subject to the Mathisson-Pirani spin supplementary condition \cite{Mathisson37, P51, Pirani56, MTW73, S99, Stephani04}\footnote{As remarked in \cite{S99}, the sign of the last term in this equation is misquoted is some references (e.g.~\cite{MTW73, Stephani04}), a problem which can be traced to the $(+--\,-)$ signature used in \cite{P51, Pirani56}.}. Notice that the total mass appearing in this equation is the sum of the Euler top's rest mass $\cM$ with its rotational kinetic energy $K$ (which is constant along the motion). The fact that the angular momentum matrix is constant on the Fermi-Walker transported frame $\{\hat{E}_1, \hat{E}_2, \hat{E}_3 \}$ is translated into the fact that the angular momentum tensor satisfies the Fermi-Walker transport equation
\[
\frac{Ds^{\alpha\beta}}{dt}=\dot{x}^\alpha \frac{D \dot{x}^\gamma}{dt} s_\gamma^{\,\,\, \beta} + \dot{x}^\beta \frac{D \dot{x}^\gamma}{dt} s^\alpha_{\,\,\, \gamma}.
\]
\end{enumerate}
\end{Remark}
%
%
%

\end{document}